\documentclass{aastex}
\usepackage{emulateapj5}
\usepackage{apjfonts}
\usepackage{epsf}
\usepackage{timesfonts}

\def\cd{{\cal D}}

\def\hb{H$\beta$}
\def\ha{H$\alpha$}
\def\mgii{Mg {\sc ii}~}
\def\civ{C {\sc iv}~}

\def\lkin{\log L_{\rm kin}}
\def\lblr{\log L_{\rm BLR}}

\def\kms{km~s$^{-1}$}

\def\mr{M_{\rm R}}
\def\mb{M_{\rm B}}

\journalinfo{The Astrophysical Journal, 615:L9-L12, 2004 November 1}

\begin{document}

\title{The Connection Between Jets, Accretion Disks, and Black Hole Mass 
in Blazars}

\author{Jian-Min Wang\altaffilmark{1},
        Bin Luo\altaffilmark{1} and
        Luis C. Ho\altaffilmark{2}}

\altaffiltext{1}{Laboratory for High Energy Astrophysics, Institute of
High Energy Physics, Chinese Academy of Science, Beijing 100039, P.~R. China.}

\altaffiltext{2}{The Observatories of the Carnegie Institution of Washington, 813
Santa Barbara Street, Pasadena, CA 91101-1292, USA}

\slugcomment{Received 2004 August 17; accepted 2004 September 22; published October 4}
\shorttitle{Jets in Blazars}
\shortauthors{WANG, LUO \& HO}

\begin{abstract}
The properties of relativistic radio jets are thought to be closely connected 
with the properties of accretion disks in active galactic nuclei.  We explore 
this issue using a sample of 35 blazars with very-long baseline observations, 
for which we can estimate the kinetic powers of their relativistic jets, the 
bolometric luminosity of their accretion disks, and the masses of their black 
holes.  Contrary to previous claims, we find that the jet kinetic power is 
significantly correlated with the disk luminosity.  This supports the notion 
that the disk is somehow coupled to the jet, even on parsec scales.  Moreover, 
we show that the correlation improves by including the black hole mass as a 
second parameter.  The dominance of the jet, parameterized as the ratio of the 
jet kinetic power to the disk luminosity, is largely controlled by and is 
inversely correlated with the Eddington ratio of the accretion disk.  This 
empirical relation should serve as a useful guide for theoretical models for 
jet formation.

\keywords{accretion, accretion disks ---  black hole physics --- 
galaxies: active --- (galaxies:) BL Lacertae objects: general --- 
galaxies: nuclei --- (galaxies:) quasars: general}
\end{abstract}

\section{Introduction}
Many of the main properties of blazars can be explained by relativistic
jets (see review by Urry \& Padovani 1995); however, jet formation remains
one of the unsolved fundamental problems in astrophysics (e.g., Meier, Koide, 
\& Uchida 2001). Three scenarios have been suggested for the origin of jets: 
(1) from the vicinity of the black hole (BH) event horizon via extraction of the  
spin energy of the BH (Blandford \& Znajek 1977); (2) from a wind from
a magnetized accretion disk (Blandford \& Payne 1982); and (3) from an 
advection-dominated accretion flow (ADAF; Narayan \& Yi 1994), either dominated 
by outflows (Blandford \& Begelman 1999) or a jet (Falcke, K\"ording, \& Markoff 
2003). Though there are still some unresolved issues with each of these models, 
it is clear that an accretion disk is involved in all of them. It is 
expected that there is a connection between the relativistic jet and the accretion 
disk.

The jet-disk connection has been extensively explored by many authors and 
in different 
ways. Since emission from the relativistic jets is highly beamed, one needs to use
isotropic radiation to investigate the connection. On large scales ($\sim$kpc--Mpc), 
many authors have tested the correlation between different emission-line luminosity 
and radio luminosity (Stockton \& MacKenty 1987; Baum \& Heckman 1989; Rawlings \& 
Saunders 1991; Falcke, Malkan, \& Biermann 1995; Xu, Livio, \& Baum 1999; Cao \& Jiang 
2001; Wang, Ho, 
\& Staubert 2003). Indeed, a strong link between the jet and the disk has been found, 
although we should note that it is indirect because the extended radio emission and 
the line emission have quite different lifetimes. On $\sim$pc scales, radio data from 
very long-baseline interferometry (VLBI) have been used to calculate the kinetic 
luminosity to test the jet-disk connection (Celotti, Padovani, \& Ghisellini 1997;
hereafter CPG). CPG found
no correlation between the broad-line region (BLR) luminosity ($L_{\rm BLR}$) and the 
kinetic luminosity ($L_{\rm kin}$), although there are some suggestive hints of a 
jet-disk connection. Maraschi \& Tavecchio (2003; hereafter MT03; see also Maraschi 
2001) investigated the problem using a sample of blazars for which the jet
power could be estimated from modeling of their spectral energy 
distribution.  They found that the jet power is linearly proportional to the 
disk power, with flat-spectrum radio quasars radiating at a greater efficiency
than BL Lac objects.  In view of the importance of this issue, it 
certainly merits reexamination.

Jet formation may depend on the state of the accretion disk.
Marscher et al. (2002) found that the radio galaxy 3C 120, as in
microquasars, ejected superluminous radio blobs following a dip
in the X-ray emission.  It is essential to probe the accretion
rates of the disks to test the jet-disk connection. Wang et
al. (2002, 2003, 2004) showed that most BL Lac objects have
optically thin ADAFs, while most radio-loud quasars have
standard accretion disk. It is interesting to note that the degree of 
radio-loudness in low-luminosity active galactic nuclei (AGNs) is determined by
the Eddington ratio (Ho 2002), and BH accreting systems
obey a ``fundamental plane'' defined by their radio emission, hard
X-ray emission, and BH mass (Merloni, Heinz, \& Di Matteo
2003). The highly anisotropic emission from the jet depends on the
dissipation of the jet kinetic energy, and the radiation
efficiency may be controlled by both the state of the central
engine and the environment of the jet (Celotti \& Fabian 1993;
Wang, Staubert, \& Ho 2002). However, how the kinetic luminosity,
a more fundamental parameter of the relativistic jet, depends on the 
properties of the accretion disk is not yet known. 

In this paper, we tackle this issue with the sample of blazars from 
CPG. We show that in blazars there is a well-defined correlation between 
kinetic luminosity, BLR luminosity, which we use as a surrogate to 
represent the bolometric luminosity of the accretion disk, and the BH
mass.  Moreover, the degree of jet-dominance diminishes with increasing 
Eddington ratio. 

\begin{table*}[t]
\begin{center}
\footnotesize
\centerline{\sc Table 1: The Sample$^a$}
\vskip 0.1cm
\begin{tabular}{cccrccllrlcl}\hline \hline
Name & Type&~~$z$ & $\cd^b$~~&$\lkin$& $\lblr$& Line& Ref.&log $M_{\rm BH}$ &$M_{\rm BH}$ Estimator& $M_{\rm B}$&Ref.\\
     &     &    &            & (erg s$^{-1}$)& (erg s$^{-1}$)&&& ($M_{\odot}$) &            & (mag) & \\
    (1)    & (2) & (3)   & (4)   & (5)    & (6)    &(7)    & (8)  &(9)~~  &(10)              &(11)&(12)\\ \hline
0336$-$019 & HPQ & 0.852 & 15.6  & 46.21  & 44.97  & \hb   & JB91 & 8.95  & FWHM(\hb)=4876 \kms   & $-$24.24  &WU02\\
0420$-$014 & HPQ & 0.915 & 16.8  & 47.07  & 45.12  & \mgii & SF97 & 8.95  & FWHM(\hb)=3000 \kms   & $-$25.84  &WU02\\
0521$-$365 & HPQ & 0.055 & 6.17  & 46.18  & 42.00  & \hb   & SF97 & 8.65  & $\sigma$=269 \kms     & ...       &WU02\\
1034$-$293 & HPQ & 0.312 & 5.4   & 45.81  & 42.80  & \hb   & SF97 & 8.66  & FWHM(\hb)=4116 \kms   & $-$23.69  &WU02\\
1253$-$055 & HPQ & 0.538 & 18    & 46.11  & 44.23  & \mgii & SF97 & 8.48  & FWHM(\hb)=3100 \kms   & $-$23.93  &WU02\\
1335$-$127 & HPQ & 0.541 & 11.9  & 46.50  & 44.15  & \mgii & SF97 & 8.42  & FWHM(\mgii)=4602 \kms & $-$22.94  &SKF93\\
1510$-$089 & HPQ & 0.361 & 14.5  & 45.48  & 44.59  & \hb   & SM87 & 8.62  & FWHM(\hb)=3180 \kms   & $-$24.41  &WU02\\
1641$+$399 & HPQ & 0.595 & 5.3   & 44.80  & 45.01  & \hb   & JB91 & 9.45  & FWHM(\hb)=5140 \kms   & $-$25.94  &WU02\\
1741$-$038 & HPQ & 1.054 & 4.3   & 46.79  & 44.44  & \mgii & C97  & 9.29  & FWHM(\mgii)=11574 \kms& $-$23.32  & F04\\
1921$-$293 & HPQ & 0.352 & 18.2  & 46.86  & 44.02  & \mgii & SF97 & 9.01  & FWHM(\hb)=9134 \kms   & $-$22.40  &JB91\\
2223$-$052 & HPQ & 1.404 & 20.9  & 46.46  & 45.28  & \mgii & SF97 & 8.81  & FWHM(\civ)=3449 \kms  & $-$25.47  &W95\\
2251$+$158 & HPQ & 0.859 & 6     & 46.19  & 45.75  & \hb   & JB91 & 9.10  & FWHM(\hb)=2800 \kms   & $-$26.64  &WU02\\
2345$-$167 & HPQ & 0.576 & 10.8  & 45.84  & 44.32  & \hb   & JB91 & 8.76  & FWHM(\hb)=4999 \kms   & $-$23.44  &WU02\\
0333$+$321 & LPQ & 1.258 & 16.6  & 47.47  & 46.30  & \mgii & C97  & 10.11 & FWHM(\mgii)=14700 \kms& $-$26.59  &SS91\\
0923$+$392 & LPQ & 0.699 & 11.6  & 46.03  & 45.23  & \hb   & O84  & 9.40  & FWHM(\hb)=7200 \kms   & $-$24.65  &WU02\\
1226$+$023 & LPQ & 0.158 & 6     & 46.09  & 45.42  & \hb   & SM87 & 9.21  &$\mb({\rm host})=-$22.11 mag        & ...       &MC04\\
1928$+$738 & LPQ & 0.302 & 4.4   & 45.53  & 44.61  & \hb   & M96  & 8.72  & FWHM(\hb)=3360 \kms   & $-$24.58  &WU02\\
1954$+$513 & LPQ & 1.220 & 6.17  & 47.24  & 44.86  & \mgii & C97  & 9.94  & FWHM(\mgii)=15241 \kms& $-$25.49  &L96\\
2134$+$004 & LPQ & 1.936 & 34.6  & 46.00  & 44.75  & \mgii & C92  & 8.91  & FWHM(\mgii)=2800 \kms & $-$27.87  &C92\\
2351$+$456 & LPQ & 2.000 & 6.17  & 46.72  & 44.76  & \mgii &SK93  & 8.80  & FWHM(\mgii)=5118 \kms & $-$24.47  &F04\\
0906$+$430 & LDQ & 0.670 & 43.4  & 46.66  & 44.67  & \mgii & C97  & 8.95  &$\mb({\rm host})=-$21.58 mag        & ...       &MC04\\
1222$+$216 & LDQ & 0.435 & 1.3   & 45.67  & 44.67  & \hb   & C97  & 8.17  & FWHM(\hb)=2197 \kms   & $-$23.92  &F04\\
1317$+$520 & LDQ & 1.060 & 1.23  & 48.62  & 46.01  & \mgii & C97  & 10.05 & FWHM(\mgii)=13900 \kms& $-$26.54  &SS91\\
1721$+$343 & LDQ & 0.206 & 1.23  & 46.97  & 44.67  & \hb   & R84  & 8.50  & FWHM(\hb)=2880 \kms   & $-$24.25  &WU02\\
1845$+$797 & LDQ & 0.057 & 1.23  & 46.31  & 43.54  & \hb   & R84  & 8.55  & FWHM(\hb)=8850 \kms   & $-$20.76  & L96\\
0235$+$164 & BL  & 0.940 & 6.5   & 45.69  & 43.81  & \mgii & C97  &$>$10.22 &$\mr({\rm host})>-$26.44 mag        & ...       & U00\\
0537$-$441 & BL  & 0.896 & 11.6  & 46.44  & 44.61  & \mgii & SF97 & 8.74  & FWHM(\mgii)=3200 \kms & $-$26.34  &W86\\
0851$+$202 & BL  & 0.306 & 8.8   & 46.37  & 43.44  & \mgii &SFK93 & 8.79  & FWHM(\hb)=3443 \kms   & $-$24.78  &SFK89\\
1101$+$384$^c$ & BL  & 0.031&3.09& 45.45  & 41.34  & \ha   & C97  & 8.29  & $\sigma$=219 \kms     & ...       &WU02\\
1308$+$326 & BL  & 0.996 & 6.8   & 45.93  & 44.63  & \mgii & SK93 & 9.24  & FWHM(\mgii)=4016 \kms & $-$27.93  &SK93\\
1538$+$149 & BL  & 0.605 & 1.3   & 44.58  & 43.05  & \mgii & SK93 & 8.94  &$\mr({\rm host})=-$23.88 mag        & ...       &U00\\
1652$+$398$^d$ & BL&0.034& 1.5   & 45.36  & 41.27  & \ha   & C97  & 9.21  & $\sigma$=372 \kms     & ...       &WU02\\
1803$+$784 & BL  & 0.684 & 8.5   & 45.74  & 44.28  & \mgii & SK93 & 8.57  & FWHM(\mgii)=2451 \kms & $-$26.66  &L96\\
1807$+$698 & BL  & 0.051 & 3.09  & 45.11  & 41.34  & \hb   & C97  & 8.52  &$\mr({\rm host})=-$23.04 mag        & ...       &F03\\
2200$+$420 & BL  & 0.069 & 4.4   & 45.13  & 42.52  & \hb   & V95  & 8.35  &$\mr({\rm host})=-$22.69 mag        & ...       &F03\\
\hline
\end{tabular}
\parbox{6.9in}
{\baselineskip 9.pt
\indent
{\sc References/Notes:--} C92: Corbin 1992;
CPG: Celotti et al. 1997;
F03: Falomo, Carangelo, \& Treves 2003;
L96: Lawrence et al. 1996;
M96: Marziani et al. 1996;
MC04: Marchesini \& Celotti 2004;
O84: Oke, Shields, \& Korycansky 1984;
SF97: Scarpa \& Falomo 1997;
SFK89: Stickel, Fried, \& K\"uhr 1989;
SKF93: Stickel, K\"uhr, \& Fired 1993;
SFK93: Stickel, Fried, \& K\"uhr 1993;
SK93: Stickel \& K\"uhr 1993;
SM:87 Stockton \& MacKenty 1987;
SS91: Steidel \& Sargent 1991;
U00: Urry et al. 2000;
V95: Vermeulen et al. 1995;
W86: Wilkes 1986;
W95: Wills et al. 1995;
WU02: Woo \& Urry 2002.

HPQ: core-dominated high-polarization quasars;
LPQ: core-dominated low-polarization quasars; 
LDQ: lobe-dominated quasars;
BL: BL Lacs.\\
$^a$All the data are scaled to luminosity distances calculated assuming
$H_0=75$ km~s$^{-1}$~Mpc$^{-1}$, $\Omega_{\rm m} = 0.3$, and
$\Omega_{\Lambda} = 0.7$.
$^b$Doppler factor.
$^c$Mrk 421 has line flux for H$\alpha$ instead of H$\beta$ or \mgii.
H$\beta$ or \mgii lines. The upper limit on its core
size given in Ghisellini et al. (1993) gives only an upper limit on 
$\lkin$. The value of $\lkin$ listed is based on the 
new VLBI observations of Piner et al. (1999).
$^d$Mrk 501 has line flux for H$\alpha$ instead of H$\beta$ or \mgii.
}
\end{center}
\vglue -0.5cm
\end{table*}
\normalsize

\section{The Sample}
For the purposes of the present paper, we need the kinetic luminosity,
bolometric luminosity, and the BH mass.
According to CPG, the kinetic luminosity is
given by $L_{\rm kin}=f\pi R^2_{\rm VLBI}n_em_ec^2\Gamma^2\beta c$, where
$R_{\rm VLBI}$ is the size of the radio jet, $\beta$ is the jet velocity in
units of the light speed ($c$), $n_e$ is the number density of relativistic
electrons, $\Gamma$ is the Lorentz factor of the jet, and $m_e$ is the
electron mass. $R_{\rm VLBI}$ is an observable parameter, while $n_e$ and
$\Gamma$ can be obtained by fitting the observed spectrum. However, the largest
source of uncertainty for the kinetic luminosity lies in the factor $f$, which
describes the particle composition of the jet.  For a jet composed of a
proton-electron plasma, $f$ has a value of $\sim 1840$, while for a
pair plasma $f =1$. As in CPG, we take $f=1$.  In the present paper,
we use a subset of the sample of CPG, which provides 
$L_{\rm BLR}$ and $L_{\rm kin}$.  Our subset is limited 
by the availability of the following data: (1) $L_{\rm kin}$, (2) BH
mass estimate, and (3) flux for the H$\beta$ $\lambda$4861 or \mgii\ 
$\lambda$2800 line, which, as described below, we use to estimate $L_{\rm BLR}$.
We omit those sources with an upper limit on $L_{\rm kin}$ because they 
only have an upper limit on $R_{\rm VLBI}$.  Ghisellini et al.
(1993) give the Doppler factors of the sample, estimated from the classical
limits of synchrotron self-Compton emission.  In this study, we estimate
BH masses using one of three methods, depending on the availability
of the necessary data: (1) the relation between stellar velocity dispersion
and BH mass, as fitted by Tremaine et al. (2002);
(2) the linewidth-luminosity-mass scaling relation, which is calibrated against
local AGNs studied using reverberation mapping (Kaspi et al.
2000); and (3) the correlation between host galaxy luminosity and BH
mass, as given by McLure \& Dunlop (2001).
When the linewidth (FWHM) of H$\beta$ is not available, we use the FWHM of \civ
or \mgii as suggested by Vestergaard (2002) and McLure \& Jarvis (2002), 
respectively, to estimate the BH mass.  Our paper analyzes 35 blazars 
(Table 1) from the sample of CPG.  
When the BH mass is estimated from the linewidth-luminosity-mass
relation, we use an extrapolation of the continuum from the magnitudes given 
in V\'eron-Cetty \& V\'eron (2003), assuming a spectrum of the form 
$F_{\nu}\propto \nu^{-0.5}$.

\begin{figure*}[t]
\centerline{\includegraphics[angle=-90,width=16cm]{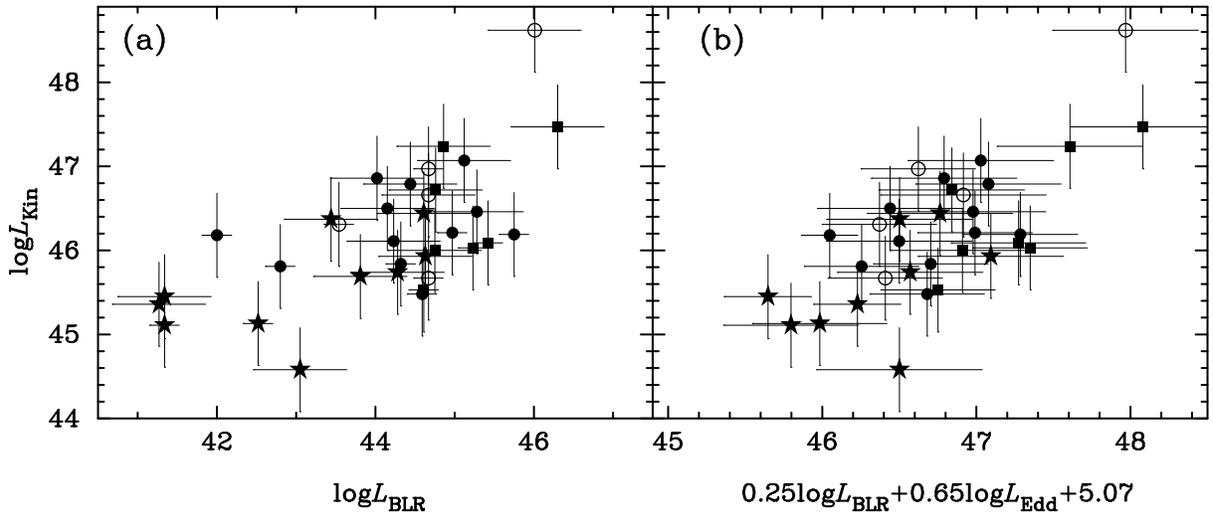}}
\figcaption{\footnotesize Correlation between jet kinetic luminosity 
and ({\it a}) $L_{\rm BLR}$ and ({\it b}) $L_{\rm BLR}$ and $L_{\rm Edd}$ for
BL Lac objects ({\it stars}), lobe-dominated quasars ({\it open circles}), 
core-dominated, low-polarization quasars ({\it squares}), and core-dominated,
high-polarization quasars ({\it filled circles}). The uncertainties on 
$L_{\rm BLR}$ are based on the standard deviation of \mgii/L$\alpha$ and 
\hb/L$\alpha$ in the composite quasar spectrum of Francis et al. (1991),
while the uncertainties on $L_{\rm kin}$ are assumed to be 0.5 dex for all 
sources. 
The error bars for the abscissa in ({\it b}) take into account the 
uncertainties in $L_{\rm BLR}$ and $M_{\rm BH}$ (see \S 2).} \label{fig1}
\end{figure*}

The uncertainty in the $M_{\rm BH}-\sigma$ relation is small, $\lesssim 0.21$
dex (Tremaine et al. 2002).  On the other hand,
the uncertainty on the zeropoint of the linewidth-luminosity-mass
relation is approximately 0.5 dex (Gebhardt et al.  2000; Ferrarese et al.
2001).  We note, however, that this method may systematically overestimate
the BH masses in radio-loud quasars if the optical-ultraviolet
continuum in these objects is significantly beamed.  Unfortunately, it is
difficult to estimate the magnitude of this effect, since the amount of
beaming is wavelength dependent (e.g., Bregman 1990), and we have no good
constraints for the optical-ultraviolet region.  According to
McLure \& Dunlop (2001), the $M_{\rm BH}-M_{\rm R}$ correlation for
quasar host galaxies has an uncertainty of 0.6 dex.  Given the
intrinsic uncertainty of the different BH mass estimators and
the heterogeneity of our sample, we estimate that the individual 
BH masses may have an uncertainty as large as $\sim$1 dex.

\section{Empirical Correlations}
The BLR luminosities given in CPG were derived by scaling several
strong emission lines to the quasar template spectrum of Francis et al. (1991),
using Ly$\alpha$ as a reference.  With this approach,  CPG find no significant
correlation between $L_{\rm BLR}$ and $L_{\rm kin}$.  We note, however, that
some of the key emission lines used in their method have a very large scatter
in their ratio with respect to Ly$\alpha$.  For example,
{\sc C iv} $\lambda$1550/Ly$\alpha$ = $63\pm 41$ and
\mgii $\lambda$2800/Ly$\alpha$ = $34\pm 20$.  The variation in the strength of
{\sc C iv} and \mgii relative to Ly$\alpha$ stems from genuine object-to-object
variations in the detailed physical conditions of the BLR in quasars.  Since
this scatter propagates directly into an error in $L_{\rm BLR}$, it is possible
that this is responsible for the apparent lack of a correlation between
$L_{\rm BLR}$ and $L_{\rm kin}$.   On the other hand, the ratio
H$\beta$/Ly$\alpha$ is largely governed by recombination physics (Netzer 1990)
and hence is much better defined.  Francis et al. (1991) find
H$\beta$/Ly$\alpha$ = $22\pm 4.1$.  Here, we will adopt H$\beta$ {\it alone} 
as the fiducial line to estimate $L_{\rm BLR}$; when H$\beta$ is not available, 
we use \mgii, which has the second smallest scatter relative to Ly$\alpha$. 
Following the notation of CPG (their Eq. 1), 
\begin{equation}
L_{\rm BLR}=\left\{
\begin{array}{l}
\frac{\langle L_{\rm BLR}^*\rangle}{L_{\rm est}(\rm H\beta)}=25.26L_{\rm H\beta},\\
  \\
\frac{\langle L_{\rm BLR}^*\rangle}{L_{\rm est}(\rm Mg II)}= 16.35L_{\rm Mg~ {\small II}},
\end{array}
\right.
\end{equation}
where $\langle L_{\rm BLR}^*\rangle=555.77$, 
$L_{\rm est}(\rm H\beta)=22$, and $L_{\rm est}(\rm Mg II)=34$. Table 1 gives
the values of $L_{\rm BLR}$ estimated from Equation 1.

Figure 1{\it a}\ shows a strong correlation between $L_{\rm BLR}$ and 
$L_{\rm kin}$.  We have tested the partial correlation among $L_{\rm kin}$, 
$L_{\rm BLR}$, and $d_{\rm L}$, the luminosity distances of the objects, so as 
to avoid the possible spurious effects caused by their common dependence on 
the luminosity distance.  Using the least square method of multivariate 
regression, we find that 
$\log L_{\rm kin}=0.43\log L_{\rm BLR}+0.11\log d_{\rm L}$, with partial
coefficient $r_{d_{\rm L}}=0.4$ and $r_{L_{\rm BLR}}=0.62$.
This test shows that the correlation between $L_{\rm kin}$ and $L_{\rm BLR}$ is 
intrinsic rather than introduced by the dependence of $d_{\rm L}$.

Since we are exploring the possible dependence of jets on the accretion disks 
(using $L_{\rm BLR}$ as a substitute for the disk bolometric luminosity), 
$L_{\rm BLR}$ should be treated as the independent variable. The least square 
method gives the following regression fit in Figure 1{\it a}\ (excluding the 
prominent outlier 3C 345):
\begin{equation}
\lkin=(0.37\pm 0.08)\lblr+(29.78\pm 3.7),
\end{equation}
with Pearson's coefficient $r=0.62$ and a probability of $p=1.0\times 10^{-4}$ 
for rejecting the null hypothesis of no correlation.  Assuming that the BLR
luminosity is due to the reprocessing of the primary continuum radiation
from the accretion disk, Equation 2 can be interpreted as evidence for a 
strong connection between the relativistic jet and the accretion disk.
The relation $L_{\rm kin}\propto L_{\rm BLR}^{0.37}$ 
is much flatter than the linear relation between jet power and disk power 
found by MT03.  It is unclear what the source of discrepancy is.  We simply 
note that our procedure for estimating disk luminosities differs from that 
of MT03.  MT03 obtain their disk luminosities either directly from the 
optical-UV luminosity of the big blue bump or from the original prescription 
of CPG.

Despite the strong correlation exhibited in Figure 1{\it a}, clearly there is 
still significant scatter.  The scatter around the best-fit line is $\sigma=0.37$ 
dex.  Although a significant fraction of this scatter may be due to 
measurement and/or systematic errors in $L_{\rm kin}$ and $L_{\rm BLR}$, it is 
possible that it indicates that a second parameter is important.  A nature 
choice to investigate is the BH mass, here cast in terms of its Eddington 
luminosity, $L_{\rm Edd}$.  Using the least square method of multivariate 
regression, we indeed find that $L_{\rm kin}$ correlates with {\it both}
$L_{\rm BLR}$ and $L_{\rm Edd}$.  Excluding 3C 345 and 0235+164 (its BH mass is
an upper limit), we obtain
\begin{equation}
\begin{array}{lll}
\lkin&=&(0.25\pm 0.09)\lblr+\\
     & &(0.65\pm 0.25)\log L_{\rm Edd}+5.07\pm 10.05,
\end{array}
\end{equation}
with Pearson's coefficient $r=0.70$. The new correlation (Fig. 1{\it b}) shows a 
smaller scatter, which has reduced to $\sigma =0.31$ dex. 
The signifcance of adding the Eddington luminosity
as a new parameter in the correlation (3) can be tested by the
$F$-test defined by Bevington (1969). We find $F_{\chi}=7.36$ and probability 
$p=1.09\times 10^{-2}$.
The improvement in the correlation by including $L_{\rm Edd}$ as an 
additional variable suggests that the jet power depends on both the 
disk luminosity (or accretion rate) {\it and}\ BH mass.  Whether additional 
variables are required, however, cannot be decided on the basis of the
data presented here. The scatter in Figure 1{\it b}, though reduced, remains 
substantial, but the significant uncertainties currently associated with the 
observables prevent us from assessing how much of the scatter is intrinsic.

Equation 3 can be casted in a different form.  Defining the ``jet-dominance'' 
factor as ${\cal F}_{\rm J}=L_{\rm kin}/L_{\rm bol}$, the Eddington ratio as
${\cal E} = L_{\rm bol}/L_{\rm Edd}$, and assuming 
$L_{\rm bol} \approx 10 L_{\rm BLR}$ (Netzer 1990), it follows that 
\begin{equation}
\log{\cal F}_{\rm J}=-0.10\log L_{\rm bol}-0.65\log{\cal E}+4.82.
\end{equation}
This implies that the relative importance of the jet power compared to the 
disk luminosity is mainly controlled by, and is inversely dependent on, the 
Eddington ratio. MT03 reached a qualitatively similar conclusion in 
their analysis.  This relation provides a physical basis for 
understanding the spectral sequence of blazars (Fossati et al. 1998;
Wang et al. 2002, 2004; MT03).
This empirical relation may also help guide theoretical models on 
jet formation.  

\section{Conclusions}
Using rough estimates of the BH masses and bolometric luminosities for a 
sample of 35 blazars with VLBI observations, we find that the kinetic power of 
the relativistic jet in these objects is strongly coupled both to the accretion 
disk luminosity and the BH mass.  Equivalently, the degree of jet dominance, 
defined as the ratio of the jet kinetic power to the bolometric luminosity of 
the disk, is largely controlled by and is inversely correlated with the 
Eddington ratio of the accretion disk.  This empirical relation should provide
strong constraints on theoretical models for jet formation and jet-disk 
coupling.

There is growing evidence in recent years that BH
accretion in AGNs provides an important source
of energy feedback to their host galaxies and their large-scale
environments (see review by Begelman 2004). Relativistic jets from
radio-loud sources, in particular, may be particularly effective
in heating the intracluster medium in galaxy clusters.  The
empirical correlations described in this study, if
they can be properly calibrated, opens up the possibility of using
them as tools to investigate quantitatively accretion-powered
feedback in the context of galaxy formation.

\acknowledgements
This research is supported by a Grant for Distinguished Young Scientists from
NSFC and 973 project.  The work of LCH is funded by the Carnegie Institution
of Washington and by NASA grants from the Space Telescope Science Institute
(operated by AURA, Inc., under NASA contract NAS5-26555).

\clearpage

\clearpage

\end{document}